# Matrix of integrated superconducting single-photon detectors with high timing resolution

Carsten Schuck[1], Wolfram H. P. Pernice[1,2], Olga Minaeva[3], Mo Li[1,4], Gregory Gol'tsman[5], Alexander V. Sergienko[3], and Hong X. Tang[1]

[1]Department of Electrical Engineering, Yale University, New Haven, CT 06511 USA
[2]Karlsruhe Institute of Technology, D-76128 Karlsruhe, Germany
[3]Department of Electrical and Computer Engineering, Boston University, Boston, MA 02215 USA
[4]Department of Electrical and Computer Engineering, University of Minnesota, Minneapolis, MN 55455 USA
[5]Department of Physics, Moscow State Pedagogical University, Moscow 119992, Russia

We demonstrate a large grid of individually addressable superconducting single photon detectors on a single chip. Each detector element is fully integrated into an independent waveguide circuit with custom functionality at telecom wavelengths. High device density is achieved by fabricating the nanowire detectors in traveling wave geometry directly on top of silicon-on-insulator waveguides. Our superconducting single-photon detector matrix includes detector designs optimized for high detection efficiency, low dark count rate and high timing accuracy. As an example we exploit the high timing resolution of a particularly short nanowire design to resolve individual photon round-trips in a cavity ring-down measurement of a silicon ring resonator.

## I. Introduction

INTEGRATED single photon detectors are key components for enabling functionality in nanophotonics and on-chip quantum optical technology. In particular quantum information processing requires efficient interfacing of photonic circuitry with single photon detectors for scalable implementations [1]. On the one hand, optical waveguide technology is one of the most promising routes to build complex quantum optical systems on-chip [2,3]. On the other hand, nanowire superconducting single photon detectors (SSPD) are emerging as the photon-counting technology best suited for integrated quantum information technology [4]. High timing accuracy, low noise and high sensitivity at telecom wavelengths show the potential to satisfy the demands of quantum technology [5].

Most of today's SSPDs are however designed for stand-alone operation and typically consist of a single detector device coupled to a single mode optical fiber [6,7]. While the compatibility of quantum waveguide circuits and SSPDs has been successfully demonstrated [8] the coupling of photons from a chip to a separate detector limits the performance of this approach. More complex [9] or larger scale [10] nanophotonic networks thus require a complementary detector architecture – ideally embedded directly into the waveguide circuitry.

Here we present a grid of hundreds of nanowire SSPDs fully integrated with nanophotonic waveguide circuits on a silicon chip. One chip contains a large number of detector designs which can all be individually addressed and characterized. On the same chip we also implement various waveguide circuits which are scalable to photonic on-chip networks. This allows us to equip each circuit with application specific SSPD designs, optimized for particular detector benchmark parameters, e.g. detection efficiency, dark count rate, timing resolution, speed, etc.

To achieve small device footprint we fabricate the optical waveguides from silicon-on-insulator substrates. The high refractive index contrast of silicon waveguides on oxide substrate layers results in strong light confinement and thus allows for very compact circuit layouts.

The SSPDs are realized as NbN nanowires patterned directly on top of the waveguides. This traveling wave design [11,12] allows for achieving very large interaction length of an incident photon with a nanowire. Traditional meander-type detectors absorb incident photons in a thin-film of a few nanometer height under normal incidence. In the travelling wave design instead, incoming light is coupled to the NbN-film along the length of the nanowire (tens of micrometers) leading to significantly increased absorption for much shorter overall wire length as compared to a meander-type SSPDs. The detection mechanism happens on picosecond timescales and is highly efficient [13]. In our case high photon absorption efficiency therefore directly translates to an increase in on-chip detection efficiency (OCDE).

Here we also show how to exploit the high timing accuracy of our detectors to observe ballistic photon transport in cavity ring-down measurements. Variable photon delay on-chip, as demonstrated here with Si-microring resonators, has interesting applications in feed-forward schemes [14] and for photon number resolving detection [15].

## II. Waveguide Integrated SSPD-Matrix

### A. Device Layout

In our layout, a single chip accommodates 240 SSPDs organized in 20 columns and 12 rows of devices which can all be addressed individually. Fig. 1a) shows an optical micrograph of a section of such a detector-grid. A single element of the SSPD-matrix used for detector characterization

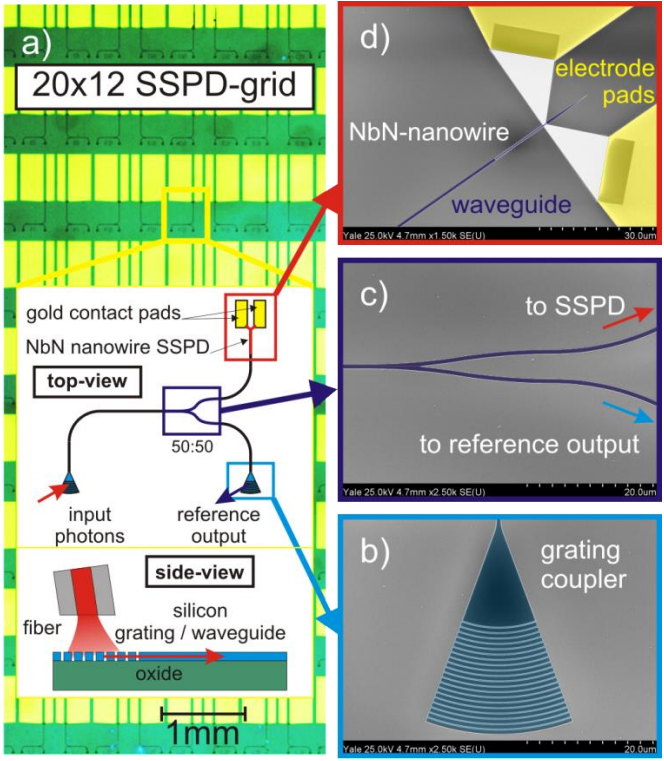

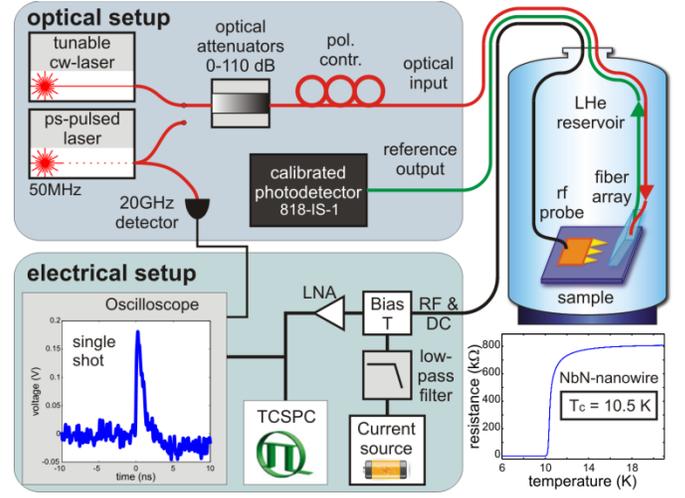

**Fig. 1** (a) SSPD grid of NbN-nanowire detectors. An individual waveguide/detector element of the matrix is shown in top and side view as an inset. It consists of three main components: (b) Grating couplers to guide light from an optical fiber into a Si-waveguide as also shown in the side-view, (c) a waveguide splitter to direct 50% of the light to the SSPD and the other half towards a reference output for detector calibration purposes, and (d) the NbN-nanowire SSPD (white) which absorbs incident photons traveling in a Si-waveguide (blue). The electrical output pulse is read out by engaging an RF probe to the electrode pads (yellow).

is shown as an inset in Fig. 1a. It consists of three main components: optical grating couplers, balanced waveguide splitters and a nanowire SSPD, all connected by low-loss waveguides.

The optical grating couplers (Fig. 1 b)) are used to couple light from a fiber array into the single-mode waveguides on the chip and vice versa. By adjusting the grating period and filling factor we optimize the coupling efficiency for a given wavelength. On each chip we include up to five different grating coupler designs for center wavelengths in the 1520-1570 nm range with a bandwidth of about 30 nm each. These grating couplers are optically reciprocal devices, i.e. the coupling loss at the input- and output ports is designed to be identical, which we confirmed on separate calibration coupler devices. To confirm correct coupler operation all devices were prescreened and discarded in case they show transmission behavior deviating significantly from the set (> 100) of reference calibration couplers. The coupler transmission of each individual device under test is then calibrated independently by using a reference output in every circuit.

**Fig. 2** Measurement setup. The sample is mounted in a liquid helium cryostat and can be optically and electrically addressed via a fiber array and an RF probe, respectively. The fabricated nanowire devices typically show a critical current of 10.5 K. Optical part: continuous wave or picosecond-pulsed laser sources (optionally) launch light via calibrated, adjustable optical attenuators into the optical input of a liquid helium cryostat. The optical output is detected with a calibrated photo-detector; electrical part: a current source (battery powered) supplies the bias current for the SSPD; the output pulses (see oscilloscope inset) are amplified with broadband, low-noise amplifiers (LNA) and registered either with a high-bandwidth oscilloscope or a time correlated single photon counting system (TCSPC, PicoQuant).

This calibration procedure is repeated for all measurement runs to precisely determine the number of photons propagating inside the waveguide towards the detector. A typical grating coupler shows a coupling loss of -13 dB at the wavelength of maximal transmission for this design.

We use a 50:50 waveguide splitter (Fig. 1c)) to route the light in equal parts from the input to the reference output and the SSPD. Such splitter devices are a standard component of the nanophotonic toolbox and yield the desired intensities at the output within 0.2 dB around 1550 nm. Both splitting ratio and splitter loss have previously been evaluated in Mach-Zehnder interferometers exhibiting interference with 33 dB extinction and no discernible loss [16,17].

The SSPD is a single U-shaped NbN-nanowire patterned directly on top of a waveguide (Fig. 1d)). Each end of the wire is connected via triangular NbN strips to large electrode pads. These pads are made of a 200 nm gold layer on top of a 5 nm Cr adhesion layer defined in an electron beam lithography step and subsequent lift-off process. Electrical contact with the electrodes is established via an RF probe to current bias the nanowire and for readout. In our SSPD-matrix we employ twelve different detector designs with varying nanowire widths (70-100 nm) and total lengths (20-80 μm). For an absorption rate of 1 dB/μm of the U-shaped NbN thin-film on top of a Si-waveguide (determined in a separate transmission measurement with 100 nm wire width) all detectors achieve long interaction length while maintaining small device footprint.

*B. Measurement Setup*

Our measurement setup is illustrated in Fig. 2. Light from a tunable laser source or alternatively from a pulsed laser source is optically attenuated to provide a pre-determined photon flux. The optical attenuators are carefully calibrated both at typical laser output levels (0-10 dBm) using a calibrated photodetector as well as at single photon levels using a single-photon detector module (id200 by IDQ). The sample is mounted inside a He4-flow cryostat, which allows for reaching temperatures down to 1.4 K. The attenuated light is coupled into the device using a single mode optical fiber array, which provides eight optical input/output ports for simultaneous excitation and readout. Light collected at the reference port is recorded with a calibrated photodetector. The number of photons arriving at the detector is then determined from the transmission through the device (grating coupler calibration at the reference port), taking into account the waveguide splitter and the external attenuation.

Electrical connection is established with a multi-contact RF probe. The nanowires are current-biased with a low-noise (battery-powered) current source. The recorded signal is amplified by two stages of electrical amplifiers and analyzed either with a time-correlated single photon counting module or a fast digital oscilloscope.

The fiber array and RF probe in our current setup allow us to simultaneously address two neighboring devices in the matrix at a time. Other devices are reached by repositioning the sample with respect to the fibers (RF probe) using low-temperature compatible translation stages. We are thus able to screen a large number of fabricated devices for their photon detection efficiency, dark count rate as well as their timing performance. Note that simultaneous operation of a larger number of devices on this chip can be achieved using existing RF-probes and fiber arrays with more elements.

Here we are mainly interested in the integration of highly efficient single photon detectors with optical waveguides. Hence, the main focus of our work lies on optimizing the coupling of light traveling inside a waveguide to the nanowire detector for efficient absorption as desired in fully integrated photonic circuit applications. The on-chip detection efficiency we measure here should therefore not be confused with the system detection efficiency usually quoted for stand-alone SSPDs. In case more efficient coupling from an optical fiber into the waveguide was desired it is possible to adapt our grating coupler design for higher efficiency (less than 1 dB loss) at the cost of a somewhat less robust fabrication procedure [18].

In our measurement configuration we do not detect appreciable levels of stray light with our detectors. To evaluate the influence of background light during measurement conditions we confirm that the detector count rate drops to the dark count level (i.e. no incident light) once the fiber array is displaced with respect to the grating couplers. Optical cross-talk between neighboring devices (250 μm separation) is analyzed by launching light into one device while monitoring the count rate of its neighboring detector which sees no input light otherwise. Again, no increase from the dark count level is observed unless the input power is increased by at least 30 dB. Stray light photons coupled into the substrate or reflected off the chip into the sample chamber only have a negligible chance to strike one of the detectors due to their tiny device footprints.

### III. DEVICE FABRICATION AND SURFACE MORPHOLOGY

Our devices are fabricated from silicon-on-insulator substrates (SOITEC) with a 110 nm silicon top layer on a buried oxide layer of 3 μm thickness. A 3.5 nm NbN thin-film is then deposited on the wafer using DC magnetron sputtering. The electrode pads and alignment marks for subsequent layers are defined in a first electron beam lithography (ebeam) step using PMMA as a lift-off resist. In a second ebeam step we employ hydrogen silsesquioxane (HSQ) resist in 3% concentration to define the nanowire detectors with high resolution. Following the development we use carefully timed reactive ion etching (RIE) in CF4 chemistry to remove the exposed NbN thin-film without attacking the silicon layer underneath. In a third and final ebeam lithography step we then pattern the waveguide circuits using HSQ in 6% concentration. The resulting resist thickness is sufficient to hold up during the subsequent RIE and inductively coupled plasma etching step in a chlorine atmosphere.

The fabricated nanowire detectors are shown in Fig. 3. High-resolution scanning electron microscopy (SEM) images reveal an alignment accuracy better than 50 nm of the NbN-nanowire (85 nm wire width shown in Fig. 3b)) on top of the 750 nm wide Si-waveguide. Note that the nanowire is buried under the HSQ masking layer which remained after etching.

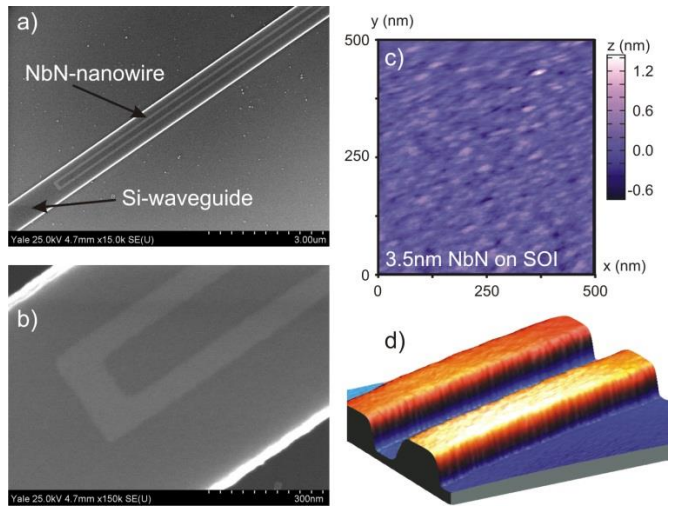

**Fig. 3** Surface morphology of a NbN-nanowire detector on top of a Si-waveguide. SEM and AFM images: a) a resist-covered NbN-nanowire of 85 nm width is visible on top of a 750 nm wide silicon waveguide; b) zoom–in of the detector region where photons are incident on the U-shaped nanowire; c) AFM scan of the deposited NbN thin-film showing RMS roughness of 1.6 Å; d) AFM image of a section of the nanowire detector protected by resist (HSQ).

We inspect a large number of devices for wire uniformity and do not find lateral defects that are significant compared to the nanowire dimensions. In order to assess the thickness uniformity of the film we perform high-resolution atomic force microscopy (AFM). As a reference we measure the surface roughness of the bare SOI-substrate and obtain an RMS value of 1.1 Å. After sputter deposition we repeat the AFM scan (see Fig. 3c)) and find an average RMS surface roughness of 1.6 Å indicating that the NbN thickness variations do not exceed a few percent. The absolute value of the NbN film thickness (3.5-4 nm) was determined from calibrated deposition rate measurements which were independently confirmed by transmission electron microscopy. Fig 3d) shows an AFM scan of a nanowire section after the second lithography step. A layer of HSQ resist protects the NbN thin film from degrading during subsequent nanofabrication steps. From the SEM and AFM inspections we conjecture that our NbN nanowires are highly uniform both in lateral and medial dimensions.

## IV. Detector Performance

An ideal nanowire SSPD should feature high detection efficiency, low noise, high speed and short timing jitter. However, often high performance in one of these disciplines comes at the cost of another. Here we have various detector geometries available within our SSPD-matrix to achieve a wide variety of performance characteristics. In this way different on-chip detector requirements can be optimally addressed by choosing the corresponding detector design: while quantum cryptography protocols require detectors with very low dark count rate, detector speed and efficiency may be a bigger concern in photon correlation experiments.

In Fig. 4 we present the results for a 70 nm wide and 80 μm long SSPD which has high performance in terms of on-chip detection efficiency reaching a maximum value of 88% (+/-5.9%) when operated at 1.64 K. The error value reflects the uncertainty of the absolute photon number arriving at the detector considering the contributions from all external and on-chip photonic components. Due to its small wire width the detector reaches high efficiency already when biased significantly below its critical current. This plateau behavior is characteristic of detectors with extremely narrow wires where the size of the hotspot originating from photon absorption approaches the nanowire width. In accordance with previous reports we here observe the onset of such a plateau at 1.64 K for telecom wavelength photons (Fig. 4). A more pronounced plateau has only been observed with ultranarrow nanowires but not for wider nanowires [19]. Note, that the high absolute on-chip detection efficiency is a result of the travelling wave design used here, allowing for very efficient absorption of photons, rather than increased quantum efficiency.

In terms of dark count rate the performance of this device is reasonable with 450 Hz at 1.4 K when biased at 99% of the critical current. Lower rates are typically found in wider and shorter wires which however did not reach equally high detection efficiency (for comparison we report a 85 nm wide, 60 μm long device with 50 Hz dark count rate reaching 55% detection efficiency; not shown). We also examine the performance close to the LHe temperature at 4 K and find the expected increase of dark count rate accompanied by a reduction of the on-chip detection efficiency to 59%.

Since this device is one of the longest ones in the matrix it has much higher kinetic inductance than shorter detectors resulting in a decay time of 1.4 ns. For high speed applications it will thus be advantageous to use the shortest detectors (20 μm in total length) in the matrix reaching 455 ps decay time.

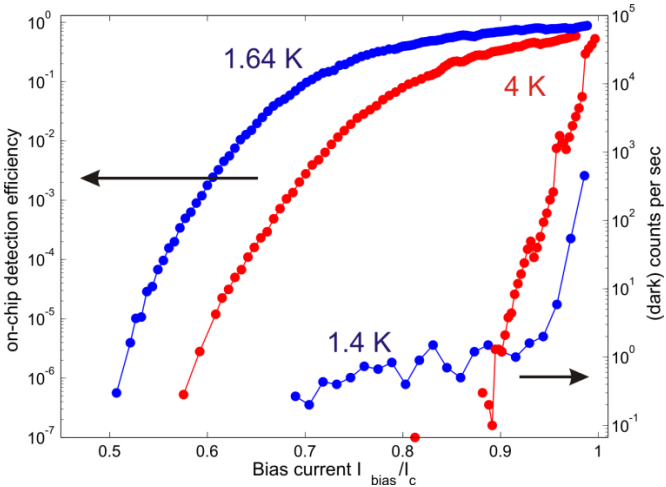

**Fig. 4** The measured detection efficiency and dark count rate for a 70nm wide and 80μm long detector device measured at 1.4K, 1.64K and 4K. At 1.64K we find a detection efficiency of 88% when the detector is biased close to Ic which decreases to 59% at 4K.

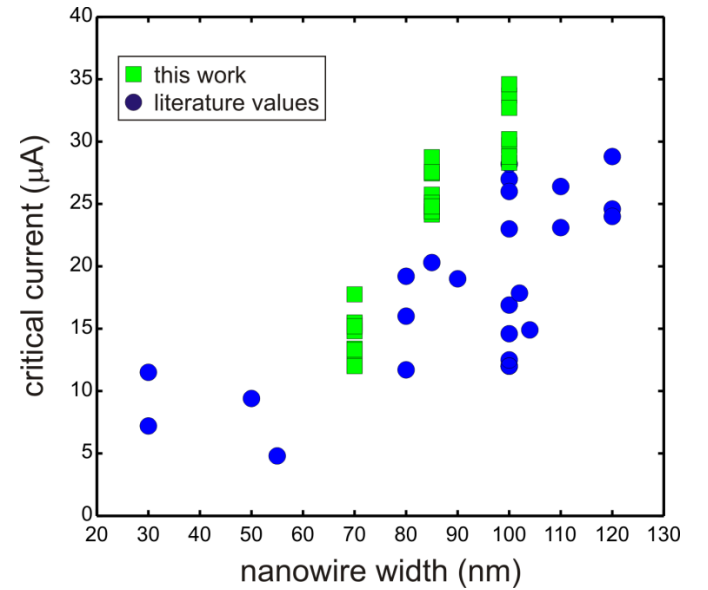

**Fig. 5** Typical critical current values measured for SSPDs with nanowire widths of 70 nm, 85 nm and 100 nm (green squares). For comparison we plot critical current values for NbN-nanowire SSPDs reported in the literature (blue dots). For similar nanowire dimensions our devices generally show higher critical current values than previously reported meander-type devices.

Furthermore, we characterize a large number of detectors in terms of critical current for the 1.4 K to 4 K temperature range. Generally high critical currents are desirable because it allows for operating the SSPD at higher bias current yielding higher output pulse amplitudes. Since the amplitude of the electrical noise is not affected by the bias current higher signal-to-noise ratios are achievable at high bias current. We estimate the critical current of nanowires with different wire widths by measuring their switching current. The results are shown in Fig. 5 where we also compare our devices to critical current values reported in the literature for state-of-the-art NbN-nanowire detectors. For our devices we observe high switching values which allow us to operate them in the high signal-to-noise regime where pulse discrimination is possible without sacrificing detection efficiency. Our results compare favorably with values measured in meander-type SSPDs which we attribute to the reduced length of our devices. The high observed critical current values also support our conjecture of high nanowire uniformity drawn in the previous section since constricted devices should switch to the normal state at lower bias currents.

Finally, the timing performance of our devices is discussed in the following section.

## V. ON-CHIP PHOTON DELAY

To demonstrate the potential of waveguide integrated SSPD grids for custom functionality we study on-chip photon delay. For this purpose our matrix includes SSPDs with particularly high timing accuracy at the output of microring resonator devices. We use these detectors to perform cavity ring-down measurements resolving individual photon round-trips. The measurement setup is shown in Fig. 6.

The microring resonators couple evanescently to the waveguide leading to the detection region. By varying the gap between waveguide and cavity the optical coupling strength can be adjusted, allowing us to operate the ring in either the under- or overcoupled regime. Typical transmission spectra measured in the through port of the device are shown in Fig.6b), illustrating the features of whispering-gallery resonances with optical quality factors on the order of a few tens of thousands. The free-spectral range of the resonator is small, because the length of the ring was chosen as 5.8 mm in order to provide a cavity roundtrip time exceeding the detector jitter.

### A. Jitter measurement

In a first step we characterize the timing performance of a 20 µm long, 100 nm wide nanowire detector in a waveguide circuit similar to the one shown in Fig. 1. In order to resolve the intrinsic detector jitter we employ a picosecond pulsed laser (Pritel), a 20 GHz bandwidth InGaAs photodetector (Agilent 83440) and a 20 GSa/s digital oscilloscope with 6 GHz real-time bandwidth (Agilent 54855A infiniium). As shown in Fig. 6, the pulsed laser output is split into two arms. The pulses in the upper arm are detected with the 20 GHz detector to provide a reference to the oscilloscope. This trigger signal is then compared to the SSPD output pulses after detecting a photon from a strongly attenuated pulse coupled into the sample chip via the lower arm. Running the

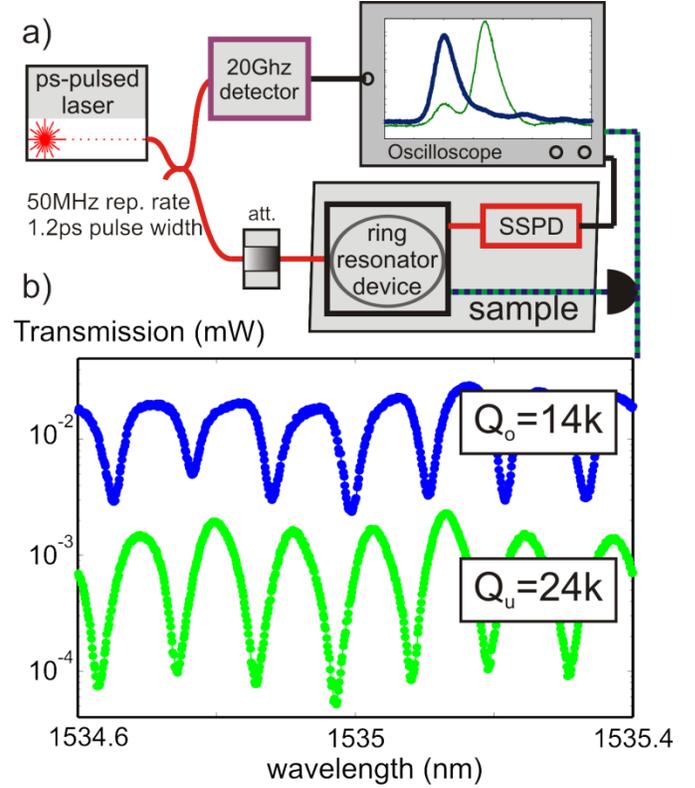

**Fig. 6** (a) Setup for the ballistic photon transport measurement with single photon detectors. (b) Transmission spectrum of undercoupled and overcoupled ring resonators.

oscilloscope in histogram mode allows for jitter measurements with true picosecond time resolution. Using the 20 GHz detector reference trigger as a start signal, the histogram is filled with stop signals triggered at the point of the maximum slope of the SSPD pulses (8 µV/ps before amplification). Using electrical amplifiers of more than 10 GHz bandwidth we find a SSPD jitter value of 18.4 ps by fitting the histogram distribution with a Gaussian function. In this case the intrinsic instrument jitter is limited by the oscilloscope bandwidth of 6 GHz.

### B. Cavity ring-down for ballistic photons

We then move on to a ring resonator device which allows us to examine its ring-down behavior in the time domain. Considering a resonator with a ring down time $\tau_0$ larger than the pulse width $T_p$ of the picosecond laser we have to treat the pulses as ballistic particles inside the resonator. Hence, the optical power circulating inside the cavity does not build up. A pulse of input intensity $I_{in}$ launched into the device will couple to the resonator and emerge in the through port as a train of pulses separated by the cavity round trip time, see Fig. 7. After passing the resonator the leading pulse will have an intensity $I_0 = t^2 I_{in}$, where $t$ is the transmission coefficient. Similarly, we can write the intensity of subsequent pulses emerging from the through port as $I_n = (1-t^2)^2 t^{2n-2} e^{-\alpha n L} I_{in}$, where $L$ is the cavity circumference and $\alpha$ describes the linear absorption inside the cavity, which is small for high optical quality factor resonators.

To achieve the $\tau_0 > T_p$ condition for the ballistic photon case, we utilize ring resonators with circumference of 5.8 mm giving rise to round-trip time of 73 ps – larger than both the detector jitter of 18.4 ps and 1.2 ps laser pulse duration.

We consider two cases, an undercoupled and an overcoupled resonator.

*C. Overcoupled resonator*

To realize the overcoupled case we launch the pulsed laser into a device with 100 nm gap between the microring and the waveguides in the through and drop ports, see Fig. 7 (top). Here the transmission coefficient is small and the majority of the light is coupled from the feeding waveguide directly into the resonator. Hence, the leading pulse emerging from the through port has smaller amplitude than the first pulse coupled out after one round-trip (Fig. 7a)).

The SSPD, with the same dimensions as the one used in the jitter measurement, is then employed to record the time-domain traces in the drop port. We bias the detector at 86% of the critical current which yields fair detection efficiency (15%) at dark count rates below 1 Hz. Fig. 7b) shows the fast decay of photons in the ring resonator which are efficiently coupled out already after just one round-trip such that only two peaks are clearly discernible. A cavity decay time of 19 ps is extracted from an exponential fit to the data in the overcoupled case, corresponding to an optical quality factor of 11,900.

For comparison the optical quality factor of the resonator is also determined from optical transmission spectra recorded at the through port using a tunable laser source. The spectrum shown in Fig. 6b) (upper blue curve) exhibits optical resonance dips separated by a small free spectral range corresponding to the large ring circumference. A Lorentzian fit yields a quality factor of 14,000 in good agreement with the value extracted from the cavity decay.

*D. Undercoupled resonator*

The undercoupled case is realized in a microring device with 200 nm coupling gap such that less light is coupled from the feeding waveguide into the ring resonator, see Fig. 7 (bottom). Otherwise the measurement configuration is the same as before (Fig. 6.). Here the transmission coefficient is large and most of the light is directly transmitted in the through port leading to large intensity of the leading pulse (Fig. 7 c)). On the contrary the light which entered the cavity now decays much slower from the cavity and we observe longer pulse trains in the drop ports as shown in Fig. 7d). The separation of the pulses again confirms the cavity round-trip time of 73 ps. An exponential fit to the four observed pulse fronts yields a cavity decay time of 38 ps for the undercoupled case, corresponding to an optical quality factor of 23,000.

For comparison we again record transmission spectra in the through port and find optical resonances with quality factors of 24,000, in good agreement with the cavity decay time.

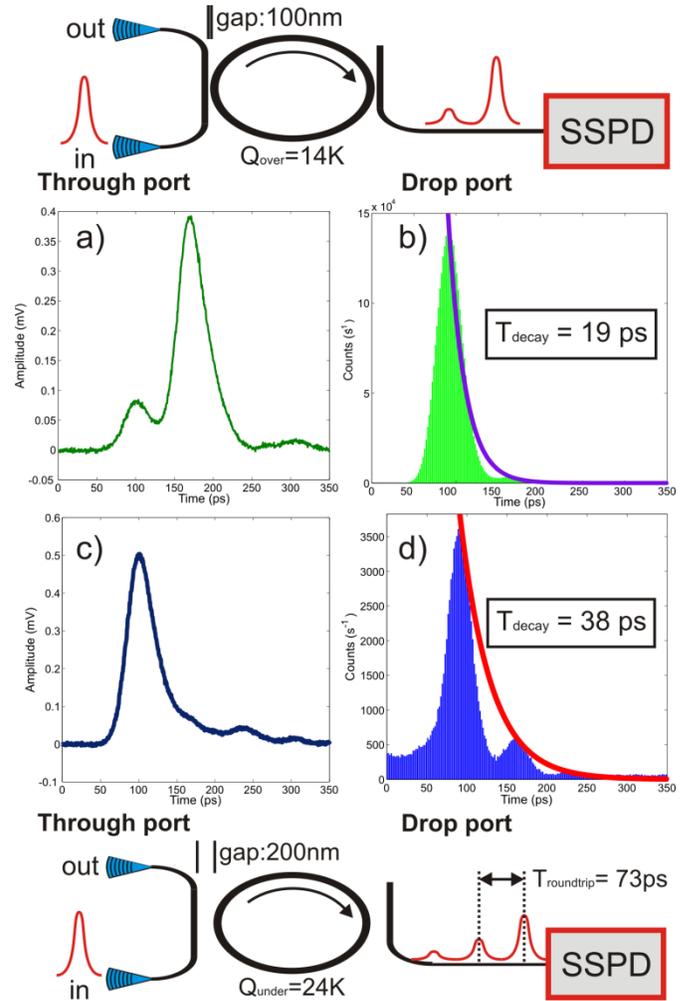

**Fig. 7** Ballistic transport measurement. Top: ring resonator in the overcoupled case (100 nm gap); a) time-domain trace of the transmission in the through port for the overcoupled case; b) time-domain trace of the photons detected by the SSPD in the drop port for the overcoupled case. An exponential fit (purple) to the data (green) yields a decay time of 19 ps; c) time-domain trace of the transmission in the through port for the undercoupled case; d) time-domain trace of the photons detected by the SSPD in the drop port for the undercoupled case. An exponential fit (red) to the data (blue) yields a decay time of 38 ps; Bottom: ring resonator in the undercoupled case (200 nm gap). The ring circumference is 5.8 mm which corresponds to a photon round trip time of 73 ps.

## VI. Conclusion

We have demonstrated a grid of SSPDs fully embedded with waveguide circuits on a silicon platform. Manufacturing waveguide integrated SSPDs in such large grids offers the possibility to use this platform as a test-bed for detector development as well as for photonic circuit characterization. Our chip contains a large variety of detector geometries and waveguide designs optimized for application specific functionality. The individually addressable detectors show optimal performance for example in terms of high on-chip detection efficiency (88%), or low dark count rate (<100 Hz), or high timing resolution (<20 ps), or combinations thereof. The integration of nanophotonic circuits with large numbers of customizable detector designs on a scalable platform will allow for satisfying many of the needs of the quantum information and photonics community [1].

As an example how a photonic circuit can be optimally characterized using a tailor-made detector design we examined time-domain multiplexing in microring resonators. We resolve individual cavity round-trips of strongly attenuated optical pulses. We expect that reducing the propagation loss [20] brings on-chip photon buffering, feed-forward schemes [14] and photon number resolving detection [15,21] within reach.